# Dimensionality-dependent type-II Weyl semimetal state in $Mo_{0.25}W_{0.75}Te_2$

Peiling Li[1†], Ya Deng[2†], Chuang-Han Hsu[3], Chao Zhu[2], Jian Cui[4], Xue Yang[1,5], Jiadong Zhou[2], Yi-Chun Hung[3], Jie Fan[1,6], Zhongqing Ji[1,6], Fanming Qu[1,5,6], Jie Shen[1], Changli Yang[1], Xiunian Jing[1,6], Hsin Lin[3], Zheng Liu[2\*], Li Lu[1,4 5,6\*], and Guangtong Liu[1,5,6\*]

[1]Beijing National Laboratory for Condensed Matter Physics, Institute of Physics, Chinese Academy of Sciences, Beijing 100190, China

[2]School of Materials Science and Engineering, Nanyang Technological University, Singapore 639798, Singapore

[3]Insitute of Physics, Academia Sinica, Taipei 115229, Taiwan

[4]Beijing Academy of Quantum Information Sciences, Beijing 100193, China

[5]University of Chinese Academy of Sciences, Beijing 100049, China

[6]Songshan Lake Materials Laboratory, Dongguan, Guangdong 523808, China

†These authors contributed equally to this work. Correspondence and requests for materials should be addressed to Z.L. (email: z.liu@ntu.edu.sg), L.L. (lilu@iphy.ac.cn), and G.L. (email: gtliu@iphy.ac.cn)

**Abstract**

Weyl nodes and Fermi arcs in type-II Weyl semimetals (WSMs) have led to lots of exotic transport phenomena. Recently, $Mo_{0.25}W_{0.75}Te_2$ has been established as a type-II WSM with Weyl points located near Fermi level, which offers an opportunity to study its intriguing band structure by electrical transport measurements. Here, by selecting a special sample with the thickness gradient across two- (2D) and three-dimensional (3D) regime, we show strong evidences that $Mo_{0.25}W_{0.75}Te_2$ is a type-II Weyl semimetal by observing the following two dimensionality-dependent transport features: 1) A chiral-anomaly-induced anisotropic magneto-conductivity enhancement, proportional to the square of in-plane magnetic field ($B_{in}^2$); 2) An additional quantum oscillation with thickness-dependent phase shift. Our theoretical calculations show that the observed quantum oscillation originates from a Weyl-orbit-like scenario due to the unique band structure of $Mo_{0.25}W_{0.75}Te_2$. The *in situ* dimensionality-tuned transport experiment offers a new strategy to search for type-II WSMs.

As a new state of matter, Weyl semimetals (WSMs)[1-3] have aroused intense interest in condensed matter physics. Together with the first WSM was experimentally discovered in TaAs, a range of novel topological properties[4-8] have been revealed, such as symmetry protected band crossing points known as Weyl nodes[4-6], topological surface states known as Fermi arcs[4-6], and anisotropic negative magneto-resistance (MR) induced by chiral anomaly[7-10]. Different from type-I WSM with a point-like Fermi surface at Weyl nodes, type-II WSMs[11-13] with tilted Weyl cones break the stringent Lorentz symmetry and exhibit lots of exotic transport properties such as Klein tunneling[14]. Recently, the $Mo_xW_{1-x}Te_2$ system has been experimentally confirmed as a type-II WSM[15-19]. Among them, our previous angle-resolved photoemission spectroscopy (ARPES) results and band structure calculations[19] have shown that $Mo_{0.25}W_{0.75}Te_2$ has larger topological Fermi arcs compared with $WTe_2$. More importantly, the Fermi energy is found to be very close to Weyl nodes[19], which makes $Mo_{0.25}W_{0.75}Te_2$ an ideal platform to observe peculiar transport properties arising from emergent Weyl fermions and corresponding topological surface states.

Chiral anomaly and Weyl orbit are two characters that researchers focus on to confirm the WSM state via transport measurements. The widely adopted method[7,8,15,16] to verify chiral anomaly at present is the detection of anisotropic negative MR. However, the negative MR could possibly be induced by other mechanisms[20-23]. For Weyl orbits, the convincing transport feature is to probe an additional quantum oscillation with thickness-dependent phase shift[16,24-26]. However, to conduct such thickness-dependent experiments, it remains challenging to control material parameters such as defects and compositions in different mechanically exfoliated sample batches. Theoretically[27,28], a WSM state only exists in 3D samples, as 2D thin samples inevitably open a gap at Weyl points and destroy the WSM state. This means that WSM states will degrade in a thinner sample. Consequently, *in situ* dimensionality-dependent magneto-transport experiments provide a new strategy to study WSM states, which allows us to investigate chiral anomaly and Weyl orbit in a single device with thickness gradient.

In this work, we consider $Mo_{0.25}W_{0.75}Te_2$ sample with thickness gradient along the *a*-axis parallel to the zigzag chain of W (or Mo). Combining low-temperature magneto-transport measurements and theoretical calculations, we demonstrate that $Mo_{0.25}W_{0.75}Te_2$ is a type-II WSM with unique surface states by observing dimensionality-dependent anisotropic negative MR and an additional quantum oscillation frequency with thickness-dependent phase shift. Our theoretical calculations suggest that this quantum oscillation is likely resulting from a scenario mimic the Weyl orbit. Moreover, the disappearance of these two characteristics in the thinner samples is consistent with our theoretical calculations. Our results contribute to a better understanding of the peculiar transport phenomena in topological WMSs.

The $Mo_{0.25}W_{0.75}Te_2$ flakes were synthesized by the molten-salt assisted CVD method[29]. More information about the sample growth is detailed in the Methods section. Figures 1a-b show the room-temperature atomic configuration and optical image of synthesized 1T′ $Mo_{0.25}W_{0.75}Te_2$ flakes. The 1T′ structure is determined by the annular dark-field scanning transmission electron microscopy (ADF-STEM) image in Supplementary Fig.

2, and the Mo-doping concentration is estimated to be around 0.2. The X-ray photoelectron spectroscopy (XPS) wide scan spectrum shown in Supplementary Fig. 3 confirms the coexistence of Mo, W and Te[30,31]. This result is also supported by the Raman spectrum in Supplementary Fig. 4, where the vibration modes of 1T′-WTe$_2$ and MoTe$_2$ coexist in the Raman spectrum of Mo$_{0.25}$W$_{0.75}$Te$_2$[30,32]. Furthermore, the peak intensity calculation from XPS spectrum in Fig. 1c indicates the composition of the synthesized alloy is stoichiometric Mo$_{0.25}$W$_{0.75}$Te$_2$. The energy-dispersive X-ray spectroscopy (EDX) shown in Fig. 1d reveals the high quality of synthesized material by observing the uniform distribution of Mo, W and Te atoms.

As mentioned above, Mo$_{0.25}$W$_{0.75}$Te$_2$ is an ideal platform for transport study on type-II WSMs. Previous studies[33,34] have shown that a structural phase transition from the 1T′ to T$_d$ phase occurs at low temperatures, which breaks the inversion symmetry and allows a type-II WSM state to emerge. Hence, we carried out low-temperature magneto-transport experiments on high-quality Mo$_{0.25}$W$_{0.75}$Te$_2$ flakes. Most importantly, the studied sample has a thickness gradient ranging from 6 to 32 nm (Supplementary Fig. 5). Compared to the bulk electron and hole mean free path $l_e$~29 nm and $l_h$~24 nm (Supplementary Table 2), the sample has both 2D-like and 3D-like transport properties. Meanwhile, the same sample guarantees that different segments have the same composition, which offers us an ideal platform to examine the exotic transport properties of a WSM in different dimensionalities. Weyl points were theoretically proposed to only exist in bulk materials[27,28], which means WSM states degrade with the sample thinned down. Thus, we can study the dimensionality effect on WSM states by simultaneously measuring the low-temperature magneto-transport properties at different segments with a thickness of $L$-nm (hereafter named Seg-$L$). The upper inset of Fig. 2a displays the temperature dependence of zero-field longitudinal resistivity $\rho_{xx}(T)$, from which one notices that Seg-28 exhibits a metallic conducting behavior. In contrast, Seg-8 shows a semiconducting behavior (upper inset of Fig. 2d), which is found to obey 2D Mott's variable range hopping (VRH)[35,36] mechanism of the conduction (Supplementary Fig. 6). Upon applying a perpendicular magnetic field to the sample, Seg-28 displays a non-saturated parabolic magneto-resistance (MR = $\frac{\rho_{xx}(B)-\rho_{xx}(0)}{\rho_{xx}(0)} \times 100\%$) of ~300% at 14 T and 0.3 K (Fig. 2a), which agrees well with previous observations for bulk WTe$_2$[37], indicating its 3D nature. The small MR upturn at low magnetic fields occurring at $T$=0.3 K is ascribed to electron-electron interactions (EEI)[38-40] as shown below. Compared with Seg-28, Seg-8 shows a small positive MR of 20% (Fig. 2d). At small magnetic fields, the steep rise of MR arises from the weak anti-localization (WAL), signifying the strong spin-orbit interaction (SOI) in the present system[14,40]. The oscillatory component superimposed on MR at high fields in Figs. 2a and 2d corresponds to quantum oscillations.

For a type-II WSM, the chiral-anomaly-induced negative longitudinal MR[15,16] is expected when the applied electric ($E$) and magnetic fields ($B$) are aligned to the tilt direction of Weyl cones. Figures 2b and 2e show in-plane magnetic field experimental results, where the field is applied along the $a$-axis of sample ($B \parallel E \parallel a$). As expected, a clear negative longitudinal MR is observed in Seg-28 in the whole measured field ranging from -14 T to 14 T. With increasing temperature, the negative MR progressively weakens and disappears around $T$=10 K. In stark contrast to negative MR in Seg-28, Seg-8 only shows positive MR. To check if the observed negative MR is induced by chiral anomaly, *in situ* tilt experiments are performed at $T$=0.3 K. The data shown in Fig. 2c illustrate that $\rho_{xx}(B)$ of Seg-28 are very sensitive to the tilt angle ($\theta$). The negative MR only occurs within $\theta \leq 15°$ and reaches a maximum at $\theta=0°$ ($B \parallel E \parallel a$). The observed negative and highly anisotropic longitudinal MR is a strong transport signature of chiral anomaly[7,8,15,16], which supports the existence of WSM states in Seg-28. Note that at small tilted magnetic fields ($|B| < 1.0$ T), $\rho_{xx}(B)$ exhibits a tilt-angle independent behavior, which can be understood by the following EEI effect[38,39] analysis. Different from Seg-28, Seg-8 displays only positive MR at all tilt angles, as shown in Fig. 2f. From the data presented above, we find that Seg-28 and Seg-8 exhibit distinct magneto-transport behaviors. This is closely related to their thicknesses, which can be interpreted as degradation of WSM states in a 2D-like region (see Supplementary Information I).

To further demonstrate chiral anomaly in $Mo_{0.25}W_{0.75}Te_2$, we perform a quantitative analysis of low-temperature magneto-transport properties and present them in Fig. 3. From Fig. 3a, we find that the zero-field resistivity $\rho_{xx}(0,T)$ from 22 K to 10 K can be well fitted by $\rho_{xx}(0,T) \propto T^3$, which is attributed to *s-d* electron scattering[41]. Below $T<10$ K, an upturn appears in $\rho_{xx}(0,T)$ curve in Fig. 3a, which can be described by EEI mechanism as demonstrated below. Considering the EEI contribution[38-40,42,43] in 2D case, the conductivity of a WSM can be expressed as (see Supplementary Information II for details),

$$\sigma_{EEI}(B,T) = \sigma_D + \frac{e^2}{2\pi^2\hbar}\left[\left(-\frac{c}{4}\tilde{F}\right) * \ln\left(\frac{k_B T\tau}{\hbar}\right) + \frac{c}{4}\tilde{F} * g_2(h)\right] \quad (1)$$

where $h = g\mu_B B/k_B T$, $g_2(h)$ has two asymptotic forms with $g_2(h) = 0.084h^2$ ($h \ll 1$) and $g_2(h) = \ln\left(\frac{h}{1.3}\right)$ ($h \gg 1$). At zero field, eq. (1) reduces into,

$$\sigma_{EEI}(0T) = \sigma_D - \frac{c}{4}\tilde{F} * \frac{e^2\ln(k_B T\tau/\hbar)}{2\pi^2\hbar} \quad (2)$$

In Fig. 3b, we plot $\sigma_{xx}(0T)$ as a function of temperature in a semi-log scale. It is found that the experimental data can be well fitted by eq. (2), confirming the low-temperature resistivity upturn arises from the EEI effect. Note that the EEI correction to the conductivity is two dimensional, which indicates that the EEI is mainly contributed by surface states.

Now we turn to discuss one of our most important findings—chiral-anomaly-induced magneto-conductivity enhancement $\sigma_{CA}$ under in-plane magnetic field ($B \parallel E \parallel a$). As shown above, the EEI also contributes a magneto-conductivity $\sigma_{EEI}$. We therefore need to remove the EEI contributions from the in-plane magneto-conductivity $\sigma_{xx}$. From Fig. 3c, one can see that $\sigma_{xx}$ can be well fitted by eq. (1) at the low-field limit ($h \ll 1$), signifying that the low-field $\sigma_{xx}$ is mainly contributed by the EEI. However, at a high-field limit ($h \gg 1$), a large discrepancy between the theoretical curves (dashed lines) and the experimental data (solid lines) indicates that there are additional contributions to $\sigma_{xx}$. To see it more clearly, we plot the experimental data of $\sigma_{xx}(14T)$ (solid blue line) and the EEI-contributed conductivity $\sigma_{EEI}(14T)$ (black dashed line) calculated with eq. (1) in Fig. 3b. It can be seen that the high-field conductivity $\sigma_{xx}(14T)$ shows an obvious enhancement compared to $\sigma_{EEI}(14T)$, highlighted by the blue shaded region. To find out the enhancement mechanism of $\sigma_{xx}$, we analyze the experimental data with the following model by considering the contributions of both the EEI and chiral anomaly,

$$\sigma_{xx} = \sigma_{EEI} + \sigma_{CA} \qquad (3)$$

According to literatures[7,8,15], $\sigma_{CA}$ is proportional to $B^2$,

$$\sigma_{CA} = C_w B^2 \qquad (4)$$

where $C_w = N \frac{e^4 v_F^3 \tau'}{4\pi^2 \hbar \Delta E^2}$ is the chiral anomaly coefficient, $N$ is the pair number of Weyl points contributing to the conductivity, $v_F$ is Fermi velocity, $\tau'$ is the axial relaxation time which can be approximately treated as intervalley scattering time[44] $\tau_i$, and $\Delta E$ is the chemical potential measured from the energy of Weyl points. We fit our experimental data with eq. (3) and plot them in the inset of Fig. 3c. The excellent agreement between the experimental data (solid yellow line) and the fitting curve (black dashed line) strongly suggests that the conductivity enhancement originates from chiral anomaly.

For a WSM, an anisotropic magneto-transport behavior originated from chiral anomaly is expected. It means that only when the external magnetic field is applied parallel to the electric field[7-10,15,16], an enhanced magneto-conductivity can be observed. Accordingly, we performed in-plane tilt experiment to further confirm our argument. Figure 3d presents magneto-conductivities measured with the in-plane magnetic field parallel ($B_{in} \parallel a$) and perpendicular ($B_{in} \parallel b$) to the current direction. Indeed, the conductivity of the former case ($B_{in} \parallel E \parallel a$) is notably larger than the latter one.

Moreover, we find that the conductivity of the latter case (light blue line) can be nicely fitted by eq. (1) (black dashed line), proving that it is completely contributed by the EEI. To quantitatively analyze the conductivity enhancement, in the inset of Fig. 3d, we plot the conductivity difference $\Delta\sigma_{xx}$ between the two directions. We find it can be well described by eq. (4), confirming that the magneto-conductivity enhancement is due to chiral anomaly. Additionally, we find that $C_w$= 48.7 T$^{-2}$·S extracted from the above fitting procedure is consistent with 44.1 T$^{-2}$·S obtained via multi-carrier fitting (Supplementary Information III) and 42.8 T$^{-2}$·S calculated from transport parameters with $N$=4 (see Supplementary Table 2). This proves that the chiral-anomaly transport involves 4 pairs of Weyl points, which is in accordance with our previous ARPES results[19]. From the above analysis based on the temperature and tilt-angle dependence of magneto-conductivity, we conclude that low-temperature conductivity enhancement originates from chiral anomaly.

Next, we move to another significant finding in our experiments—a new kind of Weyl orbit. The recent theory[25,26] and experiment[16,24,45-48] have shown that Weyl orbit can be formed by connecting surface Fermi arcs and the zeroth bulk Landau levels, which we call conventional Weyl orbit and will lead to an additional quantum oscillation in magneto-resistance curves. Moreover, the new quantum oscillation exhibits a unique thickness and tilt-angle dependent phase shift. Theoretically, the $n^{th}$ peak position of the new quantum oscillation from Weyl orbit was predicted to be[26],

$$\frac{1}{B_n} = \frac{2\pi e}{\hbar S_k}\left[(n+\gamma)\cos\theta - \frac{L}{2\pi}(\vec{k}_w \cdot \hat{B} + 2\frac{\mu}{v_\parallel})\right] \tag{5}$$

where $n$ is Landau level index, $S_k$ is the area of enclosed cyclotron orbits in $k$ space, $\gamma$ is related to Berry phase, $\theta$ is the tilt angle between $B$ and $c$-axis, $L$ is the sample thickness, $\vec{k}_w \cdot \hat{B}$ is the separation of Weyl nodes along the direction of $B$, $\mu$ is the chemical potential, and $v_\parallel$ is the Fermi velocity parallel to $B$. In our experiments, however, we find a quantum oscillation with a large $k$ space area $S_k$ (shown below), which cannot be explained by the small Fermi arcs predicted earlier[18]. Alternatively, our theoretical calculations suggest that a similar quantum oscillation could be supported by the unique band structure of $Mo_{0.25}W_{0.75}Te_2$ rather than the conventional Weyl orbit (see Figs. 4a and 4b for the comparison). As shown in Fig. 4c for 30-layer $Mo_{0.25}W_{0.75}Te_2$, the Weyl-orbit-like (WOL) orbit is composed of the surface states near the electron pocket ($S_{t2}$) of the top surface and the large Fermi arc ($S_{b1}$) of the bottom surface. The black lines and background color indicate the constant energy contours and the strength of orbital weight from the top and bottom atoms. However, when the thickness is thinned down to 10-layer, the closed orbit formed by $S_{t2}$ and $S_{b1}$ become less firmly at the transiting area marked by blue dots (see Fig. 4d), where electrons transit between $S_{t2}$ and $S_{b1}$ requiring further momentum transfer. This suggests that the WOL orbit will collapse in a thinner system and can be served as a convincing feature of the scenario.

To prove the existence of the WOL orbit, we performed *in situ* dimensionality-dependent tilt experiments. Besides Seg-28, we select other three segments in the same sample with thickness of 32 nm, 29 nm, and 26 nm (Seg-32, Seg-29, Seg-26) to conduct quantum oscillations measurements. As an example, Fig. 5a plots $d\rho_{xx}/dB$ as a function of $1/B_\perp = 1/(B\cos\theta)$ in Seg-29, where $\theta$ is the tilt angle as shown schematically in the inset of Figs. 2c and 2f. A significant feature is discerned that the oscillations depend solely on $B_\perp$, implying a 2D character. This property can be further revealed by the tilt-angle dependent oscillation frequency $F_1$, as shown in the inset of Fig. 5a (see Supplementary Fig. 8a for more data), where $F_1$ can be well described by a $1/\cos\theta$ dependence. The angle dependence suggests a 2D transport, which is associated with surface states as demonstrated below by the dimensionality-dependent experiments. Figure 5b and Supplementary Fig. 8 show the quantum oscillations $\Delta\rho_{xx}$ for five segments in the same sample with different thicknesses. The corresponding fast Fourier transform (FFT) spectra are shown in Fig. 5c, where three main oscillation frequencies $F_1 \sim 54$ T, $F_2 = 108$ T, and $F_3 = 148$ T can be seen in the 3D regime ($L \geq 26$ nm). The relatively strong amplitude of $F_2$ and $F_3$ in Seg-28 is explained in Supplementary Information IV. Interestingly, we find only $F_2' = 112$ T and $F_3' = 147$ T in Seg-8 (inset of Fig. 5d), which are very close to the value of $F_2$ and $F_3$ observed in other segments ($L \geq 26$ nm). This signifies that the corresponding oscillation frequencies ($F_2$ and $F_2'$, $F_3$ and $F_3'$) originate from the same Fermi surface. Similar value of $F_2$ and $F_3$ were previously reported in bulk $WTe_2$[49] and $Mo_xW_{1-x}Te_2$[33], and they were ascribed to bulk hole and electron pocket. Combined with our multi-carrier fitting (Supplementary Information III and Supplementary Fig. 7) and band structure calculations (Figs. 4c-d), we conclude that $F_2$ ($F_2'$) and $F_3$ ($F_3'$) come from bulk hole and electron pocket, respectively. However, the most prominent oscillation frequency $F_1$ is missed in Seg-8, indicating that it is a unique property of 3D sample. The disappearance of $F_1$ in 2D sample is consistent with our theoretical calculations that the oscillation induced by $S_{t2}$ and $S_{b1}$ will degrade in thinner samples.

Moreover, Eq. (5) tells us that an extra phase shift induced by sample thickness ($L$) and tilt-angle dependent terms ($\vec{k}_w \cdot \hat{B}$ and $2\frac{\mu}{v_\parallel}$) should be observed if Weyl orbit exists, which is distinctly different from conventional SdH oscillations[50]. Zhang[47] and Galletti[48] have recently conducted thickness-dependent experiments on Weyl orbit. However, the theoretically proposed thickness-dependent phase shift has not yet been observed. Here, we show thickness-dependent phase shift $\phi(L)$ results originating from the WOL orbit (see Supplementary Information V for discussion of tilt-angle dependent terms). From Fig. 5b, we can see the phase shift strongly depends on the sample thickness $L$, which is further corroborated by the Landau fan diagram shown in Fig. 5d. The least-square fitting (colored dashed lines) leads to the intercept of 4.29, 3.75, 3.17 for Seg-32, Seg-29, and Seg-26, respectively. These non-integer or non-half-integer phase shifts cannot be explained by only considering topological surface states with $\pi$ Berry phase[51,52]. If the phase shift comes from Weyl orbit, the sample thickness calculated with eq. (5) should be consistent with our AFM measurement result. In our case, the orbit is not formed by typical Fermi arcs, thus $\vec{k}_w$ in eq. (5) should be replaced by the vector $\vec{k}_t$ connecting two blue dots shown in Fig. 4c. By taking $\vec{k}_t \cdot \hat{B} = 0$ ($\vec{k}_t$ is in x-y plane), $\mu = 0.101$ eV, and $v_\parallel = \frac{\hbar \tilde{k}_c}{m^*} = 3.65 \times 10^5$ m/s (Supplementary Table 2 and Supplementary Information V) into eq. (5), the corresponding thicknesses are estimated to be 32 nm, 28 nm, and 24 nm for Seg-32, Seg-29, and Seg-26, respectively. The good agreement between the calculated values and AFM measurement results suggests the existence of WOL orbit. It should be noted that the different assignment of Landau index *n* may lead to different calculation results of thickness. Nevertheless, the difference of the phase shift $\Delta\phi(L)$ in the three segments is independent of *n*. The thickness difference calculated from $\Delta\phi(L)$ is 4 nm, consistent with 3 nm determined by the AFM measurement. We further calculate the frequency of Weyl orbit via the theoretical formula[26],

$$F_S = \frac{\hbar S_k}{2\pi e} \qquad (6)$$

We find the calculated value of $S_k$ from experiment results is in agreement with the theoretical value of $S'_k \sim 10^{-3}$ Å$^{-2}$, which excludes the possibility of the orbit formed by $S_{t1}$ and $S_{b1}$ (~$10^{-5}$ Å$^{-2}$). The above discussion shows that the quantum oscillation frequency $F_1$ is dimensionality-dependent, which has additional tilt-angle and thickness-dependent phase shift characteristics. All of these features are well consistent with our theoretical calculations of a new kind of Weyl orbit, which extends the generality of Weyl orbit.

In conclusion, we have prepared Mo$_{0.25}$W$_{0.75}$Te$_2$ films with thickness gradient via CVD method. Using an as-grown sample across 2D and 3D regime, our systematic low-temperature magneto-transport experiments and theoretical investigations have shown strong evidences of a type-II WSM with topological surface states. The following two characteristics, including 1) a dimensionality-dependent anisotropic magneto-conductivity enhancement proportional to $B_{in}^2$ in the configuration of $B_{in} \parallel E \parallel a$; 2) an additional dimensionality-dependent quantum oscillation with thickness-dependent phase shift, serve as convincing evidences for chiral anomaly and the new Weyl orbit, respectively. Our *in situ* dimensionality-tuned transport experiment demonstrates a new method to search for type-II WSMs, which further deepens the understanding of Fermi arc surface states and topological matters.

**Methods**

**Synthesis of atomically thin $Mo_xW_{1-x}Te_2$ alloy.** The $Mo_xW_{1-x}Te_2$ alloys were synthesized by ambient pressure CVD method.[29] The mixture of tungsten trioxide ($WO_3$) powder, molybdenum trioxide ($MoO_3$) powders and sodium chloride (NaCl) were used as the precursor and placed in a ceramic boat at the heating center. The $SiO_2$/Si substrates with polishing surface facing up were placed downstream. We used the mixed gas of $H_2$/Ar with 5/100 sccm as the carrier gas throughout the growth process. The growth temperature was found to be crucial for the composition ratio in synthesized $Mo_xW_{1-x}Te_2$ alloy. The optimized growth temperature and growth time are found to be 790 °C and 3 mins to synthesize $Mo_{0.25}W_{0.75}Te_2$.

**AFM, XPS and Raman characterization.** AFM characterizations were taken by the Asylum Research Cypher AFM in tapping mode. The XPS analysis were performed using the Kratos AXIS Supra spectrometer with a monochromatic Al K-alpha (1486.6 eV) source. Raman measurements were performed by WITEC alpha 200R Confocal Raman system.

**Devices fabrication and transport measurement.** $Mo_xW_{1-x}Te_2$ flakes were directly deposited on $SiO_2$/Si substrate, which facilitated the device fabrication without the need for transferring the materials to an insulating substrate for transport measurements. After the growth of samples, $Mo_xW_{1-x}Te_2$ alloys were firstly identified by their color contrast under optical microscopy. Then small markers were fabricated using standard e-beam lithography (EBL) near the identified sample for subsequent fabrication of Hall-bar devices. To obtain a clean interface between the electrodes and the sample, *in situ* argon plasma was employed to remove the resist residues before metal evaporation without breaking the vacuum. The Ti/Au (5/70 nm) electrodes are deposited using an electron-beam evaporator followed by standard lift-off in acetone. Transport experiments were carried out with a standard four-terminal method from room temperature to 0.3 K in a top-loading Helium-3 refrigerator with a 15 T superconducting magnet. A standard low-frequency lock-in technique was used to measure the resistance with an excitation current of 10 nA. Angular-dependent measurements were facilitated by an *in situ* home-made sample rotator.

**Computational details.** The electronic structures of $Mo_{0.25}W_{0.75}Te_2$ are calculated via the virtual crystal approximation and the applied bulk Hamiltonians of $WTe_2$ and $MoTe_2$ are obtained in the same way implemented in the previous work[18]. The finite slab Hamiltonian of $Mo_{0.25}W_{0.75}Te_2$ are extracted from the bulk Hamiltonian by truncating the real space couplings in the tight-binding scheme.

**Data availability.** The data that support the finding of this study are available from the corresponding author on request.

## Acknowledgements

We thank Xiaoxiong Liu, Zhijun Wang, and Haizhou Lu for helpful discussions. This work has been supported by the National Basic Research Program of China from the MOST under grant numbers 2016YFA0300601 and 2015CB921402, by the National Natural Science Foundation of China under grant numbers 11527806 and 11874406, by the Beijing Municipal Science & Technology Commission of China under grant number Z191100007219008, by Beijing Academy of Quantum Information Sciences under grant number Y18G08, by the Strategic Priority Research Program of the Chinese Academy of Sciences under grant number XDB33010300, by the Synergic Extreme Condition User Facility. Research in Singapore was financially supported by MOE Tier 1 grant RG4/17, MOE Tier 2 grant MOE2016-T2-1-131, and Singapore National Research Foundation under NRF award number NRF-NRFF2013-08. This research was supported by the Singapore Ministry of Education under its Tier 2 MOE2017-T2-2-136, Tier 3 MOE2018-T3-1-002 and by the National Research Foundation under its Singapore program NRF-CRP21-2018-0007, NRF-CRP22-2019-0060. H.L. acknowledges the support by the Ministry of Science and Technology (MOST) in Taiwan under grant number MOST 109-2112-M-001-014-MY3.

## Author contributions

P.L. and Y.D. contributed equally to this work. Y.D. synthesized the sample and performed XPS, EDX and Raman characterization. P.L., J.C., and X.Y. fabricated the devices and carried out the transport measurements. C.H., Y. H., and H.L. calculated and analyzed the band structures. J.Z., J.F., Z.J., F.Q., J.S., Y.H., C.Y., X.J., and L.L. discussed and commented on the manuscript. All the authors contributed to the analysis of the results and preparation of the manuscript.

**Additional information**

Supplementary information is available in the online version of the paper. Reprints and permissions information is available online at www.nature.com/reprints.

Correspondence and requests for materials should be addressed to Z.L., L.L. or G.L.

**Figure and Figure Caption**

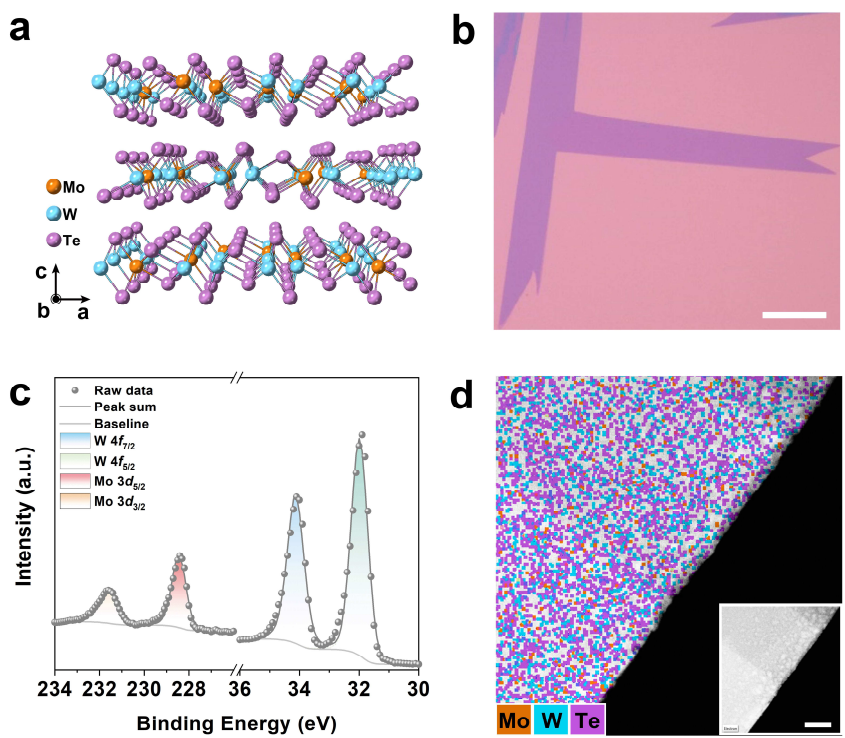

**Figure 1. Characterization of synthesized Mo$_{0.25}$W$_{0.75}$Te$_2$ flakes**. **a, b** Atomic configuration and optical image of synthesized Mo$_{0.25}$W$_{0.75}$Te$_2$ flakes. **c** The high-resolution X-ray photoelectron spectroscopy (XPS) spectrum of as-synthesized Mo$_{0.25}$W$_{0.75}$Te$_2$ flakes. The peak calculation results reveal that the composition of the synthesized alloy is stoichiometric Mo$_{0.25}$W$_{0.75}$Te$_2$. **d** The elemental energy-dispersive X-ray spectrometry (EDX) mapping of Mo$_{0.25}$W$_{0.75}$Te$_2$ flake collected from the region are shown in the inserted TEM image, revealing a uniform distribution of constituent atoms. The scale bars in **(b)** and **(d)** are 20 μm and 100 nm, respectively.

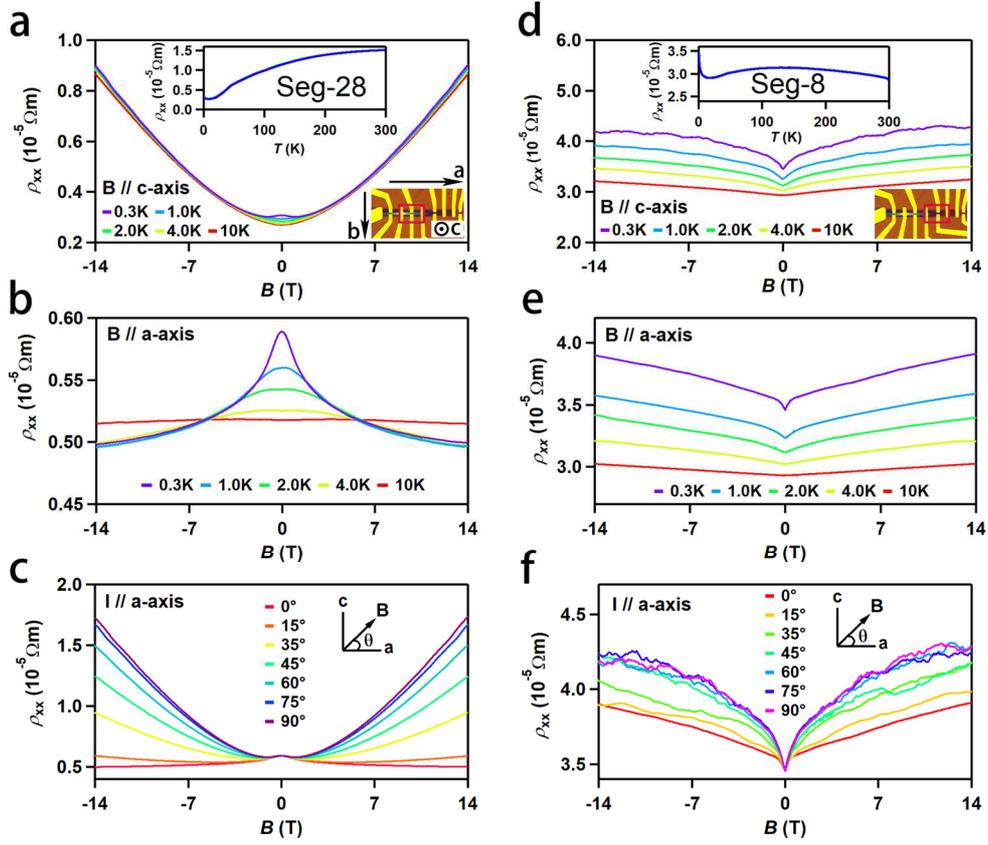

**Figure 2. Dimensionality-dependent magneto-transport characterization of the as-synthesized Mo$_{0.25}$W$_{0.75}$Te$_2$ sample. a, d** Longitudinal magneto-resistivity $\rho_{xx}$ of Seg-28 and Seg-8 in perpendicular magnetic field, respectively. The upper inset of (**a**) and (**d**) plots the zero-field resistivity of Seg-28 and Seg-8 as a function of temperatures. The region marked with solid red line in the lower right inset of (**a**) and (**d**) is the optical image of Seg-28 and Seg-8 with the sample thickness of 28 nm and 8 nm. **b, e** Longitudinal magneto-resistivity $\rho_{xx}$ of Seg-28 and Seg-8 in parallel magnetic field, respectively. **c, f** Magnetic field dependence of longitudinal magneto-resistivity $\rho_{xx}$ of Seg-28 and Seg-8 at $T = 0.3$ K with different tilt angles $\theta$. The inset is a schematic drawing of the tilt experimental setup, where $a$ and $c$ represents the crystallographic axis, $\theta$ is the tilt angle between the magnetic field $B$ and the positive direction of $c$-axis.

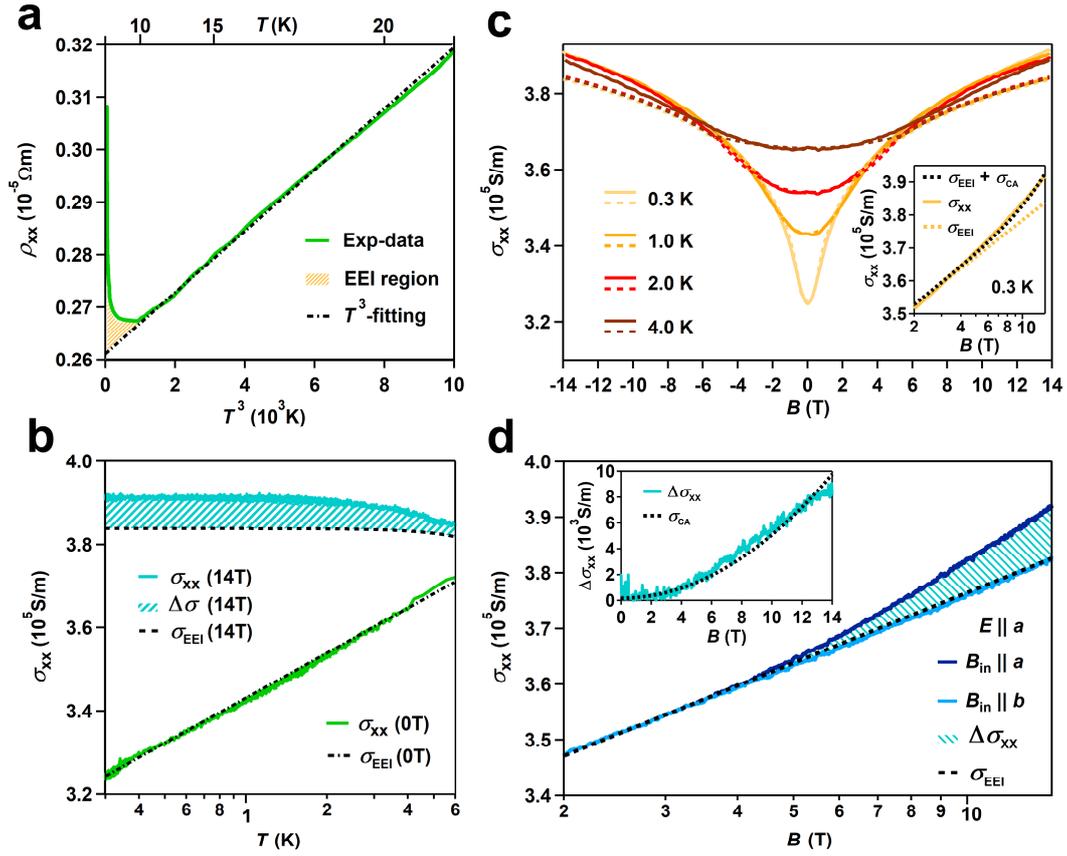

**Figure 3. Chiral-anomaly-induced magneto-conductivity enhancement in Seg-28.**
**a** The zero-field longitudinal resistivity versus $T^3$. The black dotted line represents the $T^3$-fitting curve, and the yellow shaded region indicates the EEI-induced resistivity upturn. **b** Signature of chiral-anomaly-induced magneto-conductivity enhancement. The solid blue and green lines are the temperature dependence of magneto-conductivity measured at 14 T and 0 T in-plane magnetic fields, respectively. The dashed and dotted lines are theoretical curves calculated with eqs. (1) and (2), respectively. **c** Magnetic field dependence of the in-plane $\sigma_{xx}$ collected at selected temperatures. The dashed lines are the corresponding theoretical curves calculated with eq. (1). The inset shows the magnetic field dependence of the in-plane $\sigma_{xx}$ measured at 0.3 K in a semi-log scale. The black and yellow dashed lines represent the theoretical curves calculated with eq. (5) with and without consideration of chiral anomaly contribution, respectively.

**d** Magneto-conductivities measured with the in-plane magnetic field parallel ($B_{in} \parallel a$) and perpendicular ($B_{in} \parallel b$) to the current direction. The dashed line represents the fitting results of eq. (1). The inset shows the magneto-conductivity difference between these two magnetic field directions. The agreement between the experimental data and the theoretical curve calculated with eq. (4) suggests that the magneto-conductivity enhancement is due to chiral anomaly.

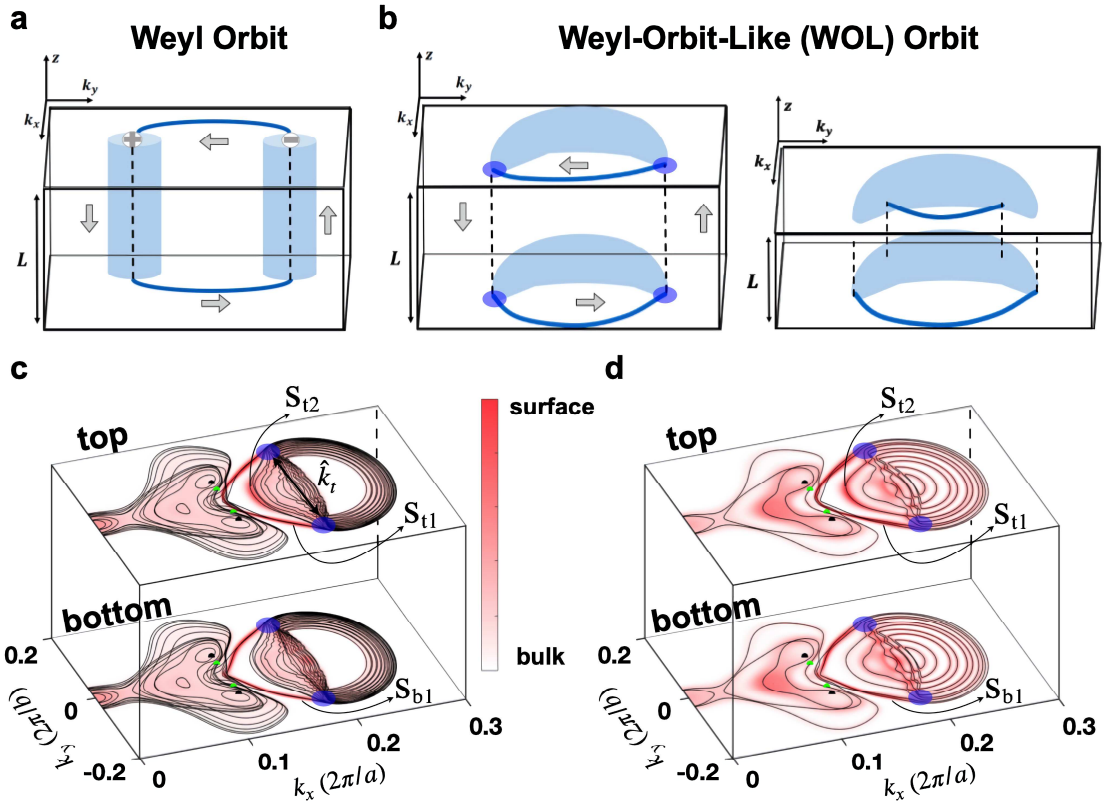

**Figure 4. Schematic figure of Weyl-Orbit-Like orbit and Fermi surface contours of 30-layer and 10-layer Mo$_{0.25}$W$_{0.75}$Te$_2$. a** The conventional Weyl orbit formed by the Fermi arcs and Landau levels of Weyl points. **b** The Weyl-Orbit-Like (WOL) orbit consists of unique surface states in a thick (left) and a thin (right) samples. Calculated Fermi surface contours and orbital weight of top and bottom atoms in **(c)** 30-layer and **(d)** 10-layer Mo$_{0.25}$W$_{0.75}$Te$_2$. In **(c, d)**, the constant energy contours and magnitude of orbital weight are indicated by the black lines and the background gradient color. Two major surface states on the top surface are indicated as S$_{t1}$ and S$_{t2}$, and the major surface states on the bottom surface are indicated as S$_{b1}$. The original Fermi arc wavevector $k_w$ considered in the conventional Weyl orbit is substituted by $k_t$ in the WOL scenario.

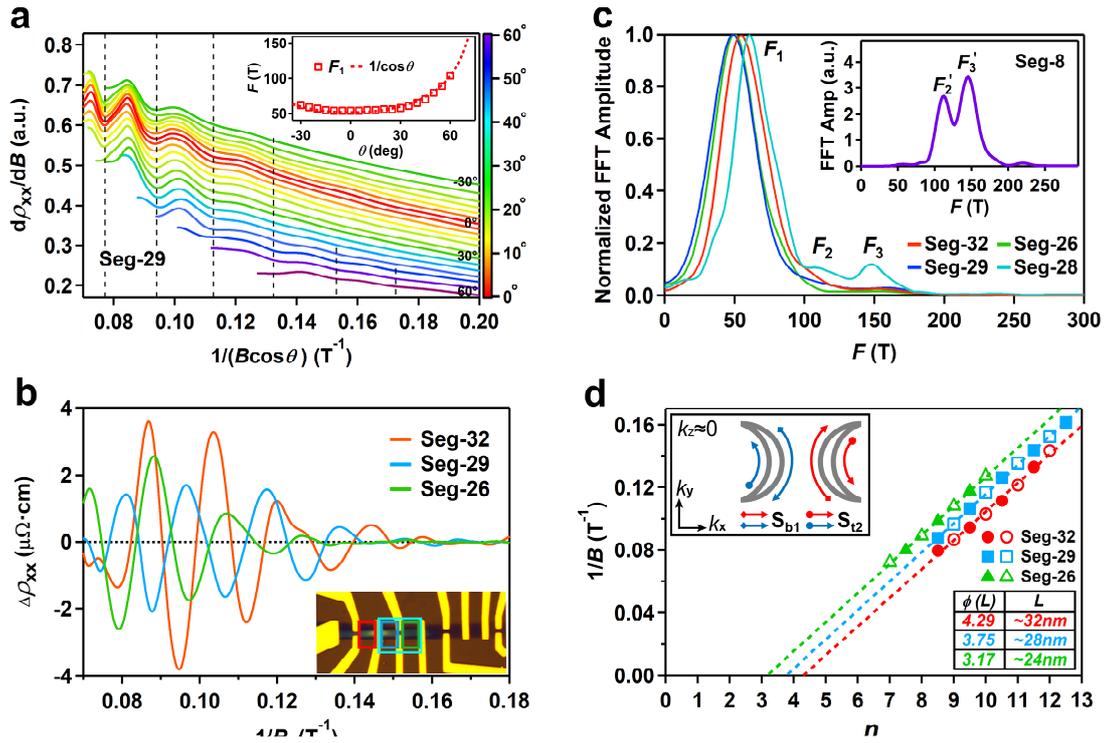

**Figure 5. Signature of Weyl orbit observed in $Mo_{0.25}W_{0.75}Te_2$ sample. a** $d\rho_{xx}/dB$ versus $1/(B\cos\theta)$ at different tilt angles. All curves are vertically shifted for clarity and curves with large $\theta(>45°)$ are horizontally shifted to compensate the additional phase shift. Vertical dashed lines emphasize the positions of minima of quantum oscillations. The inset shows the $1/\cos\theta$ dependence of the oscillation frequency. **b** The pronounced quantum oscillations $\Delta\rho_{xx}$ extracted from $\rho_{xx}$ by subtracting a smooth background for $B\|c$ ($\theta=0°$) at 0.3 K. The inset is the optical image of different segments with Seg-32 (red), Seg-29 (blue), and Seg-26 (green). Seg-28 (cyan) is composed of Seg-29 and Seg-26. **c** Fast Fourier transform (FFT) spectra obtained from quantum oscillations shown in **(b)** at different segments. Seg-28 data is calculated from oscillations shown in Supplementary Fig. 8. $F_1$, $F_2$, and $F_3$ represent three main oscillation frequencies. The inset shows FFT of quantum oscillations in Seg-8 at 0.3 K.

Only two main oscillation frequencies of $F'_2$ and $F'_3$ are observed. **d** Landau-fan diagram for quantum oscillations $\Delta\rho_{xx}(B)$ measured at 0.3 K and $\theta=0°$. The least-square fitting gives the intercepts of different segments, which are listed in the inserted table with the calculated thickness. The left upper inset shows the new Weyl orbit in $Mo_{0.25}W_{0.75}Te_2$.